\documentstyle[aaspp,12pt]{article}
\newcommand{\be} {\begin{equation}}

\newcommand{\ee} {\end{equation}}
\newfont{\mc}{cmcsc10 scaled\magstep2}
\newfont{\cmc}{cmcsc10 scaled\magstep1}
\newcommand{\co}{\rm}

\newcommand{\bc}{\begin{center}}
\newcommand{\ec}{\end{center}}
\begin{document}

\title{\large {\bf THE SOFT AND MEDIUM--ENERGY X--RAY VARIABILITY
OF NGC 5548: A REANALYSIS OF EXOSAT OBSERVATIONS}}

\author{\bf G. Tagliaferri }
\affil{Osservatorio Astronomico di Brera, V. E. Bianchi 46, 22055 Merate
(Lc), Italy}
\affil{e--mail: gtagliaf@{\co antares.merate.mi.astro.it}}

\author{\bf G. Bao\altaffilmark{1} , G. L. Israel\altaffilmark{2} }
\affil{International School for Advanced Studies (SISSA), V. Beirut 2--4,
34013 Trieste, Italy}
\affil{e--mail: Bao.gang@avh.unit.no --- gianluca@vega.sissa.it}

\author{\bf L. Stella\altaffilmark{2,3}}
\affil{Osservatorio Astronomico di Brera, V. E. Bianchi 46, 22055 Merate
(Lc), Italy}
\affil{e--mail: stella@{\co antares.merate.mi.astro.it}}

\author{\bf A. Treves }
\affil{International School for Advanced Studies (SISSA), V. Beirut 2--4,
34013 Trieste, Italy}
\affil{e--mail: treves@tsmi19.sissa.it}

\altaffiltext{1}{Current address: Physics Institute, University of
Trondheim, N--7055 Dragvoll, Norway}
\altaffiltext{2}{Affiliated to the International Center for Relativistic
Astrophysics}
\altaffiltext{3}{{\co Now at the Astronomical Observatory of Rome,
Via dell'Osservatorio 2, 00040 Monteporzio Catone, Roma}}

\begin{abstract}

We present a detailed cross--correlation (CCF) and power spectrum
re--analysis of the X--ray light curves of the bright Seyfert 1 galaxy
NGC5548 obtained with {\it EXOSAT}.  The 0.05--2 keV and 1--4 keV light
curves are cross--correlated with the 1--9 keV and 4--9 keV light curves
respectively. We discuss how spurious time lags can be introduced by
systematic effects related to detector swapping as well as the switching on
and off of the instruments. We also find strong evidence that one of the ME
detectors was not working normally during the second part of the March 1986
observation.  When these effects are taken into account, the CCF peaks are
in all cases consistent with the absence of delays between X--ray
variations at different energies. This is unlike
the results found by several authors based on the same data.

The power spectra of the 1--9 keV light curves are calculated and a
detailed search for quasi periodic oscillations (QPOs) carried out on
these spectra by using a new technique for the detection of periodic (or
quasi--periodic) signals even in the presence of source noise variability. No
significant peaks are found above the 95\% confidence detection threshold,
except during the second part of the March 1986 observation, most probably 
as a consequence of the ME detector malfunctioning.
We discuss and compare our results with those of Papadakis \& Lawrence (1993a).

\end{abstract}


\newpage

\section{Introduction}

NGC5548 is a bright, close--by (z=0.017) Seyfert 1 galaxy which was
extensively studied in different bands of the electromagnetic
spectrum. Large variability of both lines (optical--UV)
and continuum have been reported making the source an important
laboratory for exploring the mechanisms of spectral formation in AGNs. In
particular the study of correlations and time lags between the various
spectral components represent an important technique for constraining the
geometry of the emitting regions (e.g. Mushotzky et al. 1993, and
references therein). 
>From a systematic study of IUE spectra (1200--3000 \AA ~ Clavel et al. 1991)
a strong correlation between emission lines and continuum
variability was established, with
lines responding to the continuum variations with delays  of 10--70
days, depending on the degree of ionization of the species (a higher
ionization corresponds to a smaller delay).
Systematic UV--X--ray observations indicated a strong correlation in the
continuum variability in the two bands with un upper limit of $\leq$ 6 days
to any delay (Clavel et al. 1992). This, together
with the simultaneous optical UV continuum variations, showed that at least
a component of the optical--UV continuum should be generated by reprocessing
of the X--rays, rather than by the intrinsic disk variability, which should
be characterized by longer time--scales (Molendi, Maraschi \& Stella  1992).
ROSAT observations of a
soft X--ray flare with correlated variability in the UV, but without a
corresponding change in higher energy X--rays (Done et al. 1995, Walter et al.
1995) made  apparent the complexity of the  processes occurring in the
object. New simultaneous observations at various wavelengths are
currently being analysed (e.g. Korista et al. 1995 and references therein).

The importance of reprocessing on both cold and warm material
is confirmed by the observation of a Fe K
fluorescent line  at a centroid energy of $\sim$~$6.4$~keV ,  and  of a
Fe absorption edge  at $\sim$~$8$~keV
superimposed to a rather complex continuum (Nandra et al. 1991).
This  rich observational scenario  may be
accounted for by models where a hot corona above the accretion disk provides
the hard X--ray photons. At the same time these photons are in part
reprocessed by an accretion disk which in turn generates
the photons  for the Compton cooling of the electrons in the hot corona
(e.g. Haardt \& Maraschi 1993, $\dot{Z}$ycki et al. 1994).

Because of its $\sim 4$ day orbital period, allowing long uninterrupted
exposures and its wide spectral range (0.05--10 keV), the EXOSAT
satellite was particular apt to study variability and reprocessing in the
X--ray band on time scales of tens of hours or less. For a systematic study
of the variability of Seyferts and QSO contained in the EXOSAT database we
refer to Grandi et al. (1992) and Green, Mc Hardy \& Lehto (1993).

NGC 5548 was observed with EXOSAT 12 times in 1984--86 with a total exposure
of $\sim$200 hrs. Flux variations up to a factor of 4 and 3
between the various observations and within each observation, respectively,
were clearly seen. A detailed analysis of the EXOSAT light curves
was performed by several authors.
Variability on  timescales of hours was studied by Kaastra
\& Barr (1989, hereafter KB), who searched for delays between variations
in the soft (0.05--2 keV) and medium (2--6 keV) energy X--ray light curves,
suggesting that soft X--rays variations lead by $\sim 4600\pm 1200$ {\it s}.
This result, if confirmed,
would have profound theoretical implications, favoring models, such as those
mentioned above, where the medium energy X--rays are produced by Compton
scattering off the UV/soft X--ray photons by electrons in a hot corona.
Walter and Courvoisier (1990) re--examined the same data by using a
different analysis technique substantially confirming the results of KB.

Papadakis \& Lawrence (1993a, hereafter PL) performed a power spectrum
analysis the EXOSAT ME light curves and reported the likely detection of
quasi--periodic oscillations (QPOs) in 5 out of 8 observations.
They suggested that the frequency ($\sim$2~mHz) of the QPOs
increases with source intensity, whereas the fractional root mean square
amplitude of variability decreases as the source brightens.
Since this behaviour is similar to that of e.g. compact galactic
X--ray binaries, PL suggest that intensity--correlated
QPOs in NGC5548 may also arise from instability or variability in an
accretion disk around a massive black hole.

Because of the quality of the EXOSAT light curves and the importance of
the physical inferences derived from them, we feel justified
in presenting a new and independent analysis of relatively old data.
This is done in the light of
some systematic uncertainties which may arise in the analysis of the EXOSAT
light curves from relatively faint sources (such as most AGNs). These
uncertainties are discussed in greater detail in a paper on the BL Lac object
PKS 2155--304 (Tagliaferri et al. 1991, hereafter Paper I).

In Section 2.1 we summarise the  characteristics of the EXOSAT light curves of
NGC5548.  Our CCF analysis for different X--ray energy bands is described in
Section 2.2. Details on our search for
QPOs are given in section 2.3. The conclusions are in Section 3.

\section{Data Analysis }

\subsection{Light Curves}

The data considered here have been obtained through the
EXOSAT database  and refer to
the low energy imaging telescope (LE) and the medium energy experiment (ME)
(White \& Peacock 1988).
The Argon chambers of the ME experiment consisted of an array of 8
collimated proportional counters mainly sensitive to 1--20 keV X--rays.
In order to monitor the background
rates and spectrum the ME
was generally operated with half of the detector array pointed at the
target, and half at a nearby source--free region.
The two halves were usually interchanged every 3--4 hours, a procedure
indicated as {\it array swap} (hereafter {\it AS}).
The LE telescope was used with a Channel Multiplier Array (CMA)
in the focal plane. The CMA was sensitive to the 0.05--2.0 keV energy band
and had no intrinsic energy resolution (De Korte et al. 1981); however, a
set of filters with different spectral transmission could be interposed in
front of the detector.

The EXOSAT observations of NGC 5548 considered here
are summarised in Table 1.
Column 1 gives a letter identifying each
observation, column 2 the observing date, column 3 the ME exposure time,
column 4 the number of {\it AS},
{\co column 5 the r.m.s. dispersion calculated from the 1-9 keV light curves
with a binning time of 1000 s, column 6 the expected r.m.s. dispersion from
counting statistics}
and column 7 the average 1--9 keV ME count
rate (per ME array half). Columns 8 through 10
give the exposure times in the LE telescope used in conjunction with
the Aluminium/Parylene (Al/P),
Boron (Bor) and thin--Lexan (3Lex) filters, respectively.
The ME data products (energy spectra and light curves)
stored in the EXOSAT database have been
given a quality flag ranging from 0 (unusable) to 5 (excellent). Data
with quality flag between 3 and 5 are of sufficiently good quality
for a detailed analysis (see {\it The EXOSAT Database System:
available databases} (1991)). In our analysis we have considered
all observations with the ME quality factor $\ge 3$; this excludes
two observations carried out in March 1984 and January 1985
(not listed in Table 1) with exposure times of $\sim 36000$ and
$\sim 22000$ s, respectively.
An example of light curves is shown in Fig. 1.

\subsection{Cross Correlation Analysis}

In order to study the possible delays between the intensity variations in
the various bands we calculated the CCF of the LE (0.05--2 keV) and ME
light curves (1--9 keV); this is indicated as LE/ME.  For the LE we
considered only the light curves that were obtained with the 3Lex filter
(which provided the highest photon throughput) and were longer than $\sim
15000$~s. This allows to search for delays longer than one hour.
Observations I and J  are the only suitable for the LE/ME analysis.  We
have also cross--correlated the 1--4 keV to the 4--9 keV ME light curves
(ME/ME): This subdivision of the ME range for NGC 5548 provides
comparable count rates in the two bands. The ME/ME CCF
analysis was also performed for all the other observations given in Table 1.
We used the standard CCF algorithm contained in the
timing analysis package {\it Xronos}
(Stella \& Angelini 1992), which is well suited for
equispaced and (nearly) continuous data, such as the EXOSAT light curves
of NGC~5548.

In the analysis of the EXOSAT light curves of PKS 2155--304 (Paper I) we
identified a number of systematic effects that may alter the CCF.  These
are briefly summarized here. A possible problem is related to the {\it AS}
procedure, which leaves an uncertainty of up to $\pm 0.5 \ cts \ s^{-1}$
in the level of the background subtraction. If the results of the CCF
analysis change significantly by adding or subtracting one of the light
curves across the {\it AS} with a constant value of up to $0.5 \ cts \
s^{-1}$, then these results should be considered with
\setcounter{footnote}{0}
caution.\footnote{\co Due to a misprint, paper I reports an uncertainty of
up to $\pm 0.05 \ cts \ s^{-1}$ the correct value is the one reported
here (Parmar \& Izzo 1986; A.N. Parmar, private communication).}
Moreover,
the presence of {\it AS} implies  that the ME light curves are interrupted
by gaps of typical duration of 15 minutes. Although these durations are
short compared to the entire light curves, the discontinuity
that they introduce in the ME light curves can have strong effects on the
CCF.  To reduce the effects of the gaps, we fill them with the running
average of the light curve calculated over a duration of $\sim 1.5$ hour.
With this choice the moving average follows the light curve behaviour on
time scales of hours, while the statistical fluctuations are reduced
due to the relatively high number of points used in the average. For a given
observation, the start and end times of the LE and ME light curves usually
differ from a few minutes to tens of minutes.  This can alter the shape of
the CCF, especially if standard algorithms are used as in our case (see
Paper I). To avoid the problem altogether one should therefore make sure
that the two light curves are strictly simultaneous, disregarding the data
intervals in which only one light curve is available.

In our analysis we rebinned the light curves in time bins of 1000~s.
We excluded all bins with an exposure time of less than 50\% .
In some cases we used also intensity windows in order to exclude
those bins in the original light curves (resolution of 4--10 s) which
were clearly affected by an inadequate background subtraction.

The CCF analysis of observation J, the longest and that for which KB
report delays between soft and hard variations, was performed in two
different ways both for the ME/LE and ME/ME cases. First we considered the
entire LE and ME light curves (see Fig.~1), with the gaps bridged with the
running mean 
{\co and did not exclude the non-simultaneous parts of the light curves.}
%
A $\sim 6$~hr
interruption is apparent close to the beginning of the observation, due
to the switch off of the EXOSAT instruments at the perigee passage
(see Fig.~1). If the $\sim 1.5$~hr
long light curve interval that preceeds this long gap, is included
in the analysis, than the LE/ME CCF shows a marked asymmetric peak centered
around a delay of $\sim 7000-8000$ (Fig.~2).
{\co If we exclude the gap or impose the simultaneity of the two light curves,
then the delay is much shorter or not present at all. 
We note that the light curve shown in Fig.~2 of KB paper 
not only includes the gap, but in the LE it includes
another 30 ks of data at the end of the NGC~5548 observation, when this
source was seen serendipitously in the field of the BL Lac object
1E1415.6-2557 (Giommi et al. 1987). Of course there are no ME data
for this additional interval. It appears that for their timing analysis KB
used all the data shown in their Fig.~2, altough this
is not explicitely stated. As we have shown the non simultaneity of the two
data sets strongly affects the results of CCF analysis (see also paper I).
Therefore, we did not consider the extra LE data on NGC~5548 during the 
EXOSAT observation of 1E1415.6-2557. Moreover, in the rest of all our analysis
we considered only strictly simultaneous data.}
We then {\co also} excluded the first $\sim7.5$~hr
from our analysis of the light curves of observation J.
The LE/ME and ME/ME CCFs calculated in this way are shown in Fig. 3.
A peak is clearly present in the CCFs which is in both cases
asymmetric and centered near zero time delay. To derive quantitative
information on possible delays between the variations in the soft and hard
X--ray light curves, we fitted the central peak of the CCF with a Gaussian
function plus a constant. 
{\co In the case of the ME/ME CCF, a linear term was added to the fit,
to account for the stronger asymmetry of the CCF (see Fig. 3).}
The results for centroid of the peak are
$+400$~s (90\% confidence interval: $-800 \ +1500$~s) and
$+1500$~s ($+500 \ +2700$~s) respectively.

The second procedure consisted in dividing the light curve into three
segments, the first two segments, about 6 and 7 hour long, containing only
one {\it AS}, and the third segment, of about 11 hours, containing two
{\it AS}. The CCF was calculated for each segment. This treatment reduced
the possible effects of the {\it AS} discontinuity on the CCF; however it
has the disadvantage of decreasing the longest detectable delay time
(about 2--3 hours, a value still consistent with the delay time reported
by KB).  
There is essentially no peak in the second intervals both for the LE/ME
and ME/ME CCFs, 
{\co while in the first interval in both CCFs there is a weak peak centered
on zero time delay}
(Figs 4a--d). In the third segment a
clear peak is present in the ME/ME CCF, while a feature with a negative
and a positive component can be noted in the LE/ME CCF (Figs 4e,f).  This
feature is due to the presence of the two {\it AS}, indeed if we
consider only one of the two {\it AS} each time, then the first {\it AS}
gives rise only the negative peak, whereas the second {\it AS} causes only
the positive peak.  Moreover, by looking at the ME light curves it seems
that the central parts (between the two {\it AS}, see fig 5) are not
properly aligned with the other two. We tested how stable the CCF peaks
are to the addition of a constant value to the central parts. For instance
in the ME/ME CCF the peak disappears completely by adding 0.1 and 0.4
cts~s$^{-1}$ to the central 1--4 and 4--9 keV light curves respectively (Figs
5a,b shows the two ME light curves before and after having added the above
constant values to the central part). As a further test we performed
an LE/ME and an ME/ME CCF analysis by considering only the first 13~hrs of
this observations (i.e. 3 {\it AS}). Again no peak is present in either CCF.
We can conclude that the segmented analysis does
not provide evidence for delays between the LE/ME and ME/ME variations.

Another problem emerged through a careful inspection of the ME light
curves (Figs 1 and 5). One can see that the light curve intervals between
the fourth and fifth {\it AS} and after the sixth {\it AS} are much
noisier than the others (this is seen even more clearly in light curves
with a somewhat shorter binning time). This is probably due to one of the
three aligned detectors (first half of the ME in this case) not behaving
normally. That this behaviour arises from one of the detectors
(and not from the source) is confirmed
{\co both by the lac of it in the LE light curve and by the fact that 
between the fifth and sixth array swap, when the relevant half of the ME
array is offset, one of the detectors (detector B as reported in the ``ME
Obervation Log book" A. Parmar, private comunication) was switched off, due
to malfunctioning. After the sixth array swap detector B was switched on
again, but it was clearly not functioning properly, yet (see Figs 1 and 5).}
%
%
In this case the malfunctioning detector should be excluded from the
analysis, something that was not done in the automatic analysis that
generated the ME database products for this observation. 
We conclude that detector B in the first ME half is most likely
responsible for the extra variability in the ME light curves. We
intended to repeat the analysis  starting from the ME raw data.
However we could not obtain the original data from ESA, since the
relevant magnetic tape turned out to be unreadable (A. Parmar,
private communication).

For all other observations in Table 1, because of the shorter exposure
times and therefore lower number of {\it AS} (see Tab. 1), we considered
the cross correlation of the entire light curves with the data gaps
bridged by the running mean. For observation I, the second longest,
the LE/ME CCF is again flat, while the ME/ME CCF shows a strong peak
centered around zero time delay (Fig 6). It can be seen from the figure,
however, that the {\it half width at half maximum} of this peak is
comparable to the duration of the light curve segments between
{\it AS}; this suggests that the peak might be due to the systematic
uncertainties in the ME background subtraction across the {\it AS}.
To test the reliability of this CCF peak we subtracted 0.2 $cts~s^{-1}$
to the first part of both ME (1--4 and 4--9 keV)
light curves (before the first {\it AS}) and added 0.2 $cts~s^{-1}$ to
the second and third parts of the 4--9 keV light curve, trying to reduce the
discontinuity due to {\it AS} visible in the light curves. Again this was
sufficient to make the CCF peak disappear. Various other tests showed that
by adding or subtracting 0.1--0.2 $cts~s^{-1}$ (which are well within the
systematic uncertainties of the detector background subtraction) to
selected segments of the ME light curves in between {\it AS}, the peak
can become more pronounced or disappear altogether.
%
{\co Also the r.m.s. dispersion is clearly affected by 
the {\it AS}; indeed if we add 0.3 and 0.5
$cts~s^{-1}$ to the 1-9 keV light curve before the first and after the
last {\it AS}, the resulting r.m.s. dispersion is 0.28 to be
compared with the value of 0.39 given in Table 1.}

For all other observations, we calculated only the ME/ME CCF.
No peak was detected in the CCFs of observations A, C, D, E, F and H,
{\co
consistent with the fact that no significant variability is present
in the 1-9 keV light curves (see Table 1).}
Instead a clear peak was detected in observations B and G (Figs 7,8).
In both cases the peak is not centered around zero delay. This would indicate
that the variations in the 1--4 keV light curves precede the variations in
the 4--9 keV light curve by about 2000--3000~s. However, we believe that
also these delays are spurious. In the case of observation G the peak is
almost certainly due to background subtraction uncertainties across the
{\it AS}. Fig. 9a gives the original ME light curves, while Fig. 9b shows the
same light curves after the addition of a constant value of 0.2 cts~s$^{-1}$
to the segments before the {\it AS}. The latter light curves show virtually
no discontinuity across the {\it AS} and the resulting CCF is flat.
%
{\co Again by adding 0.5 $cts~s^{-1}$ to the 1-9 keV light curve before the
{\it AS}, the resulting r.m.s. dispersion is 0.18 to be
compared with 0.34 of Table 1.
We also used observation G to test whether the abrupt discontinuity
introduced by the {\it AS} can cause the asymmetry seen in
some of our CCFs. For instance, the CCF peak in Fig. 8
is steeper on the right hand side. If we subtract a
constant from both ME light curves before the {\it AS} (increasing
the {\it AS} discontinuity, see Fig. 9a) the peak asymmetry becames
more prounanced. Instead by subtracting 0.5 $cts~s^{-1}$ from both ME
light curves {\it after} the {\it AS} (changing the discontinuity from a
step-up to a step-down), then an asymmetric peak, which is steeper on the
left hand side, is obtained. This clearly shows that the
discontinuities introduced by the
{\it AS} procedure can also make the shape of the CCF peak asymmetric.}
%
%
For observation B, that has no {\it AS}, the peak is probably due
to instability in the background which are then reflected in the
background--subtracted source light curves. Indeed if we cross--correlate
either one of the source light curves in the two energy bands with the
light curve of the background we find
{\co a negative peak centered around zero time delay,
which indicates an excess of background subtraction.}
%

\subsection{Search for QPOs}

We re--analysed the 1--9~keV ME light curves from the
observations in Table~1, in order to carry out a detailed
search for the QPOs with frequencies  of $\sim 1-2.5 \times 10^{-3}$~Hz
reported by PL. The 120~s binned light curve from each observation
was divided in M consecutive intervals of $\sim 1-2$~hr duration and
the average power spectrum calculated over the power spectra from individual
intervals. This allowed to approximately reproduce the frequency range
and resolution used by PL in their analysis.
Values of M equal to 23, 17, 8, {\co 10 and 17} were used for observations 
J, C, I, A--F--G and B--D, respectively.
This method of analysis reduces by about one decade the low frequency end
of the power spectra, such that only marginal evidence is found
for the increase towards low frequencies, that reflects the {\it red noise}
variability of the source. In any case,
to search for QPOs we adopted a recently developed
technique to detect significant power spectrum peaks even in the
presence of ``coloured" noise components arising from the source variability
(Israel \& Stella 1995; Stella et al. 1995). 
The technique relies upon a suitable
smoothing algorithm in order to model the continuum power spectrum
components underlying any possible peak.
By dividing the power spectrum by the smoothed spectrum, a flat (white
noise--like) spectrum is produced, the statistical properties of which
are worked out as the ratio of two random variables of know distribution,
namely the power spectrum and the smoothed spectrum.
A search for oscillations is then carried out  by looking
for peaks in the divided power spectrum which exceed a given
detection threshold.

Selected average spectra and the
corresponding $95\%$ confidence detection thresholds
are shown in Fig~10. No significant peaks exceeding the
threshold  
were found in the frequency range $\sim$4$\cdot$10$^{-4}$~--4$\cdot$
10$^{-3}$ Hz for any of the power spectra from observations C, {\co B--D, 
A--F--G}, J and I.
Observation J was also analysed in different time intervals,
in consideration of the possible malfunctioning of one of the ME detectors
during the second half of the observation (see Section 2.2).
This was done by calculating a power spectrum for the source light
curve and a power spectrum from the corresponding background light curve
during each of the 4 array swap--free intervals in between the third array swap
and the  end of the observations. 
These power spectra and the corresponding
$95 \%$ confidence detection thresholds are shown in Fig~11. Significant
peaks are clearly detected in the second and fourth power spectra from the
source at a frequency of about $1.8$ and $2.6 \times 10^{-3}$~Hz,
respectively. It is very likely that these peaks were caused by 
some kind of quasi--periodic
instability in the detector of the first half of the ME array that did not
function properly during the second half of observation J.
The LE light curves (0.05--2.0 keV), characterized by a
poorer signal to noise ratio, were also searched for
QPOs; only negative results were found.

Our results argue against the detection of QPOs in the X--ray flux of
NGC~5548 reported by PL. The power spectrum technique used by PL
involves averaging the logarithm of the power spectra from different intervals
therefore producing power estimates that approximately follow a Gaussian
distribution (Papadakis \& Lawrence 1993b). Model fitting can then be
performed using standard least square techniques. The continuum power
spectrum components are well fitted by a constant (representing the
counting statistics noise) plus a power law (describing the source red
noise). According to PL the grouped power spectra from the three longest 
observations (C, I and J) display  a $95\%$ significant QPO peak (as estimated
through an F--test after the addition of a Guassian to the model function).
However, we have shown that the QPO during observation J very likely
arise from a detector problem.

PL devised also a test to evaluate the significance of power spectrum
peaks from individual observations. The best fit model
(a power law plus a constant) is used to estimate the continuum
components. The power spectrum is then divided by the best fit model in
order to produce a white noise power spectrum in which the presence of
statistically significant peaks is tested. PL found, in 3 out of 5 cases,
a peak in the $1.1-2.4 \times 10^{-3}$~Hz frequency range at a significance 
level of $>95\%$. However, PL did not take into account the statistical 
uncertainties introduced in the divided power spectrum by the uncertainties 
in the best fit model (as evidenced by the lack of any mention of them), 
therefore overestimating the significance of the peaks.
To reassess this significance, we extracted the power spectra from
Fig.~1 {\co of PL} fitted them with a constant, after excluding the
power estimates corresponding to the peaks and the red noise.
These constants together with their $1\sigma$ uncertainties on these 
averages, were then used to work
out the distribution of the divided spectrum in a way that parallels
the method of Israel \& Stella (1995). Based on this distribution the
significance of the peaks in Fig.~2 of PL was evaluated again. The divided
power specta of observations A--F--G (G3 in Table 1 of PL) and observation 
I are characterised by a peak with a significance of $\sim 95 \%$ and
$\sim 88\%$, respectively. These values are lower than those worked out
by PL ($\sim 98 \%$ and  $\sim 96\%$, respectively).
The power spectrum of observation J, which
formally contains the most significant peak, was disregarded in
consideration of the detector problem discussed above.

\section{Conclusion}

Our re--analysis of the CCFs of EXOSAT ME light curves of NGC5548
%
{\co does not confirm}
%
the claim of KB of a $\sim 5000$~s delay between the
medium and soft X--rays variations. This was considered as a strong
argument in favour of models where medium energy X--rays are produced by
scattering of softer photons. Our results do not exclude this
possibility. We note however that the 1990 ROSAT observations
(Nandra et al. 1993) detected a variability pattern hardly consistent with
very soft X--ray variations (0.1--0.4 keV) preceding
the variations of somewhat harder X--rays
(1--2.5 keV). This indicates the complexity of physical processes
occurring in the source.

Our power spectrum analysis does not confirm the detection of
QPOs in the mHz range reported by PL. In particular we have shown that the
only power spectrum peak with a significance of $>95\%$
most probably results from the malfunctioning of one ME detector.
The argument of PL according to which the black hole mass of NGC 5548 has
an embarrassingly low value of a few hundred thousand solar masses,
loses its validity.

While the results of this paper are essentially ``negative", we hope that
our work contribute illustrating subtle effects which may yield spurious
results in the analysis of X--ray light curves from AGNs.

\acknowledgments{We thank an anonymous referee for her/his very usefull
comments and suggestions.}
\newpage

\begin{table}[htb]
\begin{center}
\begin{tabular}{lcccccllll}
\multicolumn{8}{c}{\bf Table 1}\\
\tableline \tableline
Obs & Date & Me exp. &{\it rms} disp. &expe. {\it rms} &No. of &Counts &Al/P&Bor&3Lex\\
&  &(s)      &$cts~s^{-1}$ & $cts~s^{-1}$& AS & $cts~s^{-1}$&(s)&(s)&(s)\\
\tableline
A & 84/032  &17010  & 0.23 & 0.27 &0  &3.32$\pm$0.18 &1902 &7153  &3202 \\
B & 84/062  &32630  & 0.26 & 0.20 &0  &4.44$\pm$0.04 &3483 &3679  &2763 \\
C & 84/193  &59430  & 0.26 & 0.27 &3  &2.88$\pm$0.03 &11872&26488 &11053\\
D & 85/062  &26800  & 0.22 & 0.20 &1  &3.65$\pm$0.04 &4434 &11413 &4605 \\
E & 85/159  &25750  & 0.30 & 0.26 &1  &1.43$\pm$0.05 &6675 &9915  &3824 \\
F & 85/173  &17020  & 0.30 & 0.20 &2  &3.03$\pm$0.05 &4252 &6672  &3311 \\
G & 85/186 &23370   & 0.34 & 0.20 &1  &1.82$\pm$0.04 &3613 &10963 &3095 \\
H & 85/195  &19020  & 0.22 & 0.19 &1  &1.35$\pm$0.05 &4040 &      &2622 \\
I & 86/019  &59860  & 0.39 & 0.22 &3  &4.97$\pm$0.02 &     &3469  &39037\\
J & 86/062  &83830  & 0.45 & 0.29 &6  &3.82$\pm$0.02 &2392 &4017  &69809\\
\tableline
\end{tabular}
\end{center}
\end{table}

\newpage
\section*{ Figure captions}

{\bf Figure~1:} NGC5548 LE (0.05--2 keV) and ME (1--4 and 4--9 keV) light
curves during the longest EXOSAT observation (1986/062, observation J through
the paper). The arrows show the array swaps of the ME detector halves,
the corrspective data gaps of about 15 minutes are filled with the running
mean (see text).
Note the big gap at the beginning due to the switch off of the
detectors at the satellite perigee passage.

\vspace{4mm}

{\co {\bf Figure~2:} LE/ME light curve cross correlations of observation J.
All data shown in Fig. 1 has been used. Note that the peak centered 
around a delay of $\sim 7-8$ ks is also clearly asymmetric.} 

\vspace{4mm}

{\bf Figure~3:} LE/ME (panel a) and ME/ME (panel b) light curve
cross correlations of observation J. A clear peak around zero time
lag is clearly present in both cases. 
{\co The Gaussian plus constant model fit to the central peak is also
shown. The fit was carried out over a range of lags of $\pm 40$ ks and
$\pm 20$ ks respectively. In the ME/ME case, a linear term was
added to the fit, in order to account for the CCF asymmetry.}

\vspace{4mm}

{\bf Figure~4:} cross correlations of the light curves of
observation J divided in three different segments (see text).
A clear peak is present only in the third part (panels e--f).

\vspace{4mm}

{\bf Figure~5:} top panel: final part of the ME light curves of observation J,
the arrows show the array swaps of the ME detector halves. Note the
noisier light curves before and after the first and last array swap.
Bottom panel the same light curves after having added 0.1 and 0.4
$cts~s^{-1}$ to the central 1--4 and 4--9 keV light curves respectively;
the discontinuity due to the detector array swaps is clearly reduced.

\vspace{4mm}

{\bf Figure~6:} cross correlations of the ME light curves of
observation I. Again a strong peak centered on zero delay is
clearly present.

\vspace{4mm}

{\bf Figure~7:} cross correlations of the ME light curves of
observation B. The peak is not consistent with zero delay time, and
would imply that the variations in the 1--4 keV light curve precede
the variations in the 4--9 keV light curve. However this result
is probably spurious, see text.

\vspace{4mm}

{\bf Figure~8:} cross correlations of the ME light curves of
observation G. The peak is not consistent with zero delay time, and
would imply that the variations in the 1--4 keV light curve precede
the variations in the 4--9 keV light curve. However this result
is probably spurious, see text.

\vspace{4mm}

{\bf Figure~9:} top panel: ME light curves of observation G, the
arrows show the array swap of the ME detector halves. Bottom panel:
the same light curves after having added 0.2 $cts~s^{-1}$ to the
two light curves before the array swap; the discontinuity due
to the detector array swap is clearly reduced.

\vspace{4mm}

{\bf Figure~10:} Power spectra from the EXOSAT ME 1--9 keV light curves of
observations {\co A--F--G (84/032, 85/173 and 85/186), B--D (84/062 and 
85/062), C (84/193), and I (86/019)} (from top to bottom); the solid lines 
give the corresponding 95\% confidence detection thresholds.

\vspace{4mm}

{\bf Figure~11:} EXOSAT ME 1--9 keV Power spectra of the ME
light curves of NGC~5548 (left) and the background (right) during
the second half of observation J. Each panel refers to an array
swap free interval, starting from the third array swap (see text
for details). The ME array half used in each panel is indicated.
The solid lines give the 95\% confidence detection thresholds.

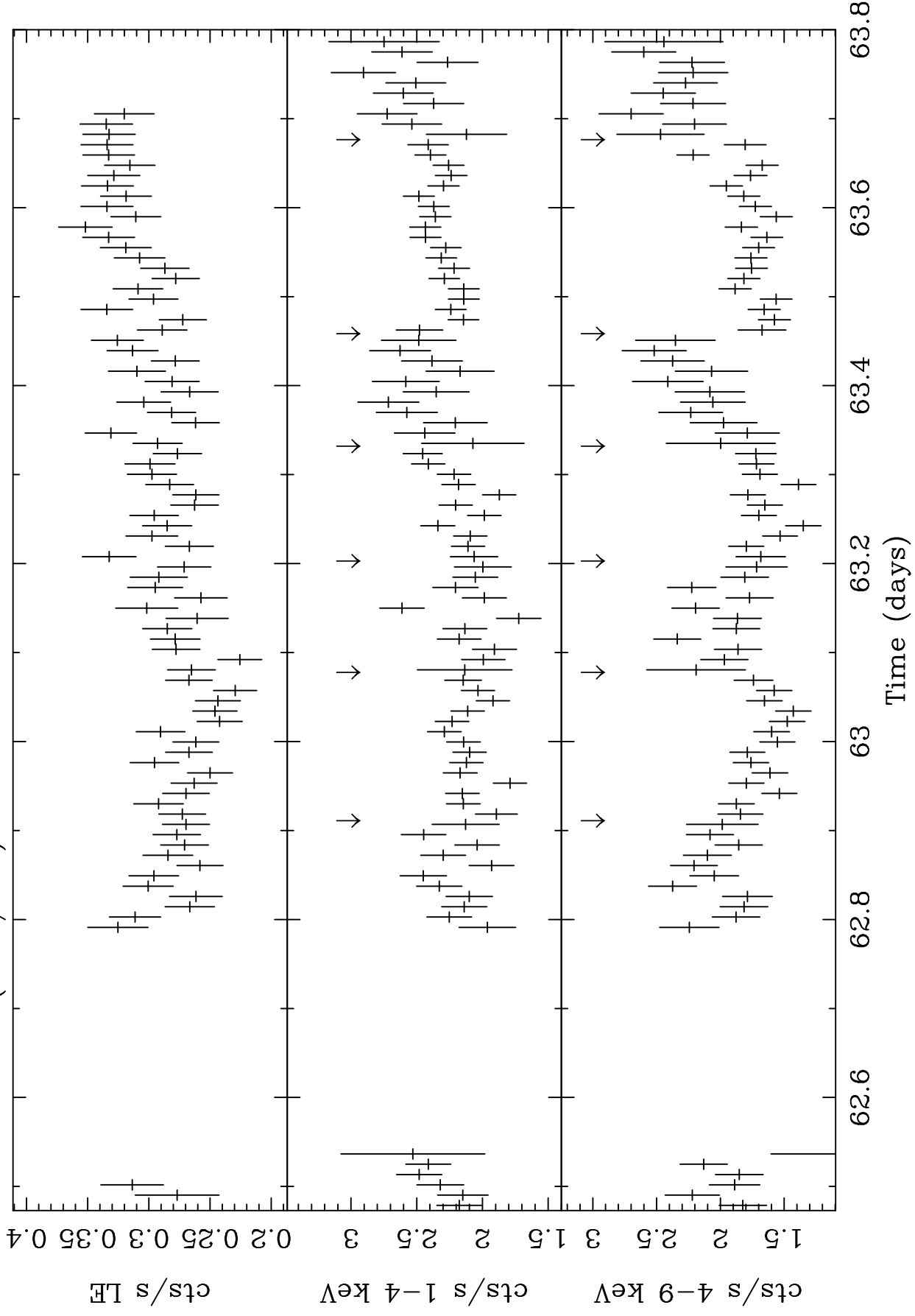
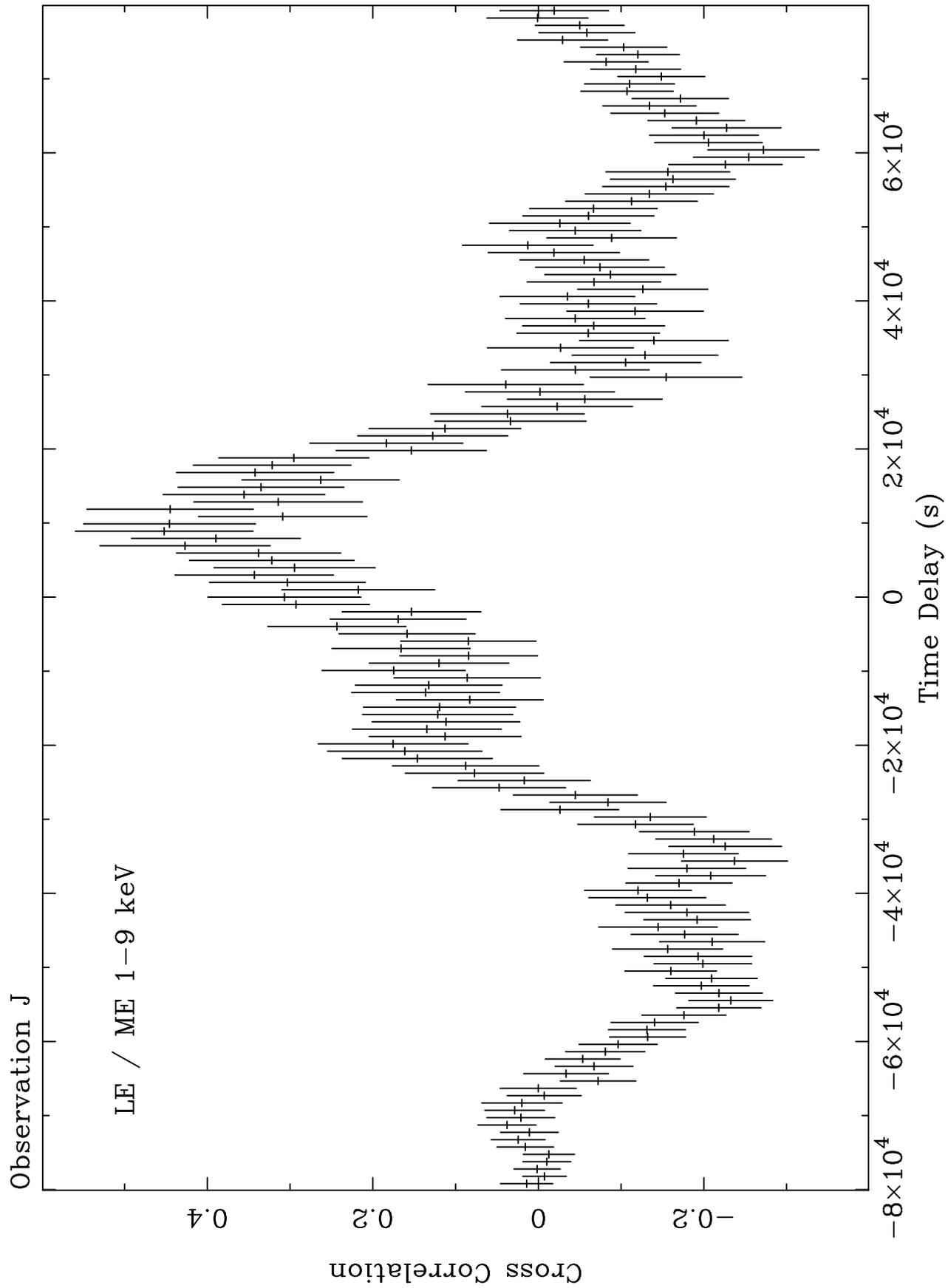
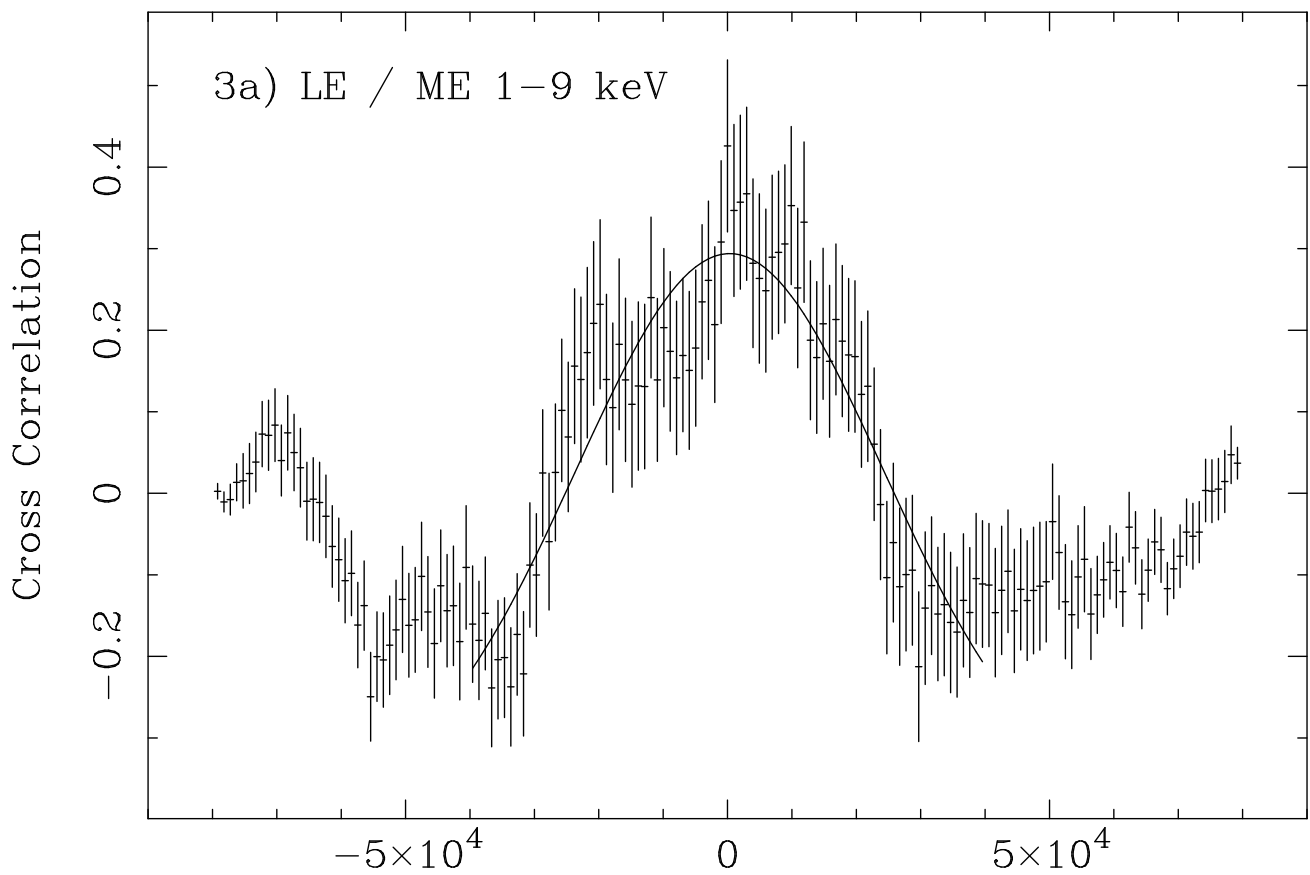
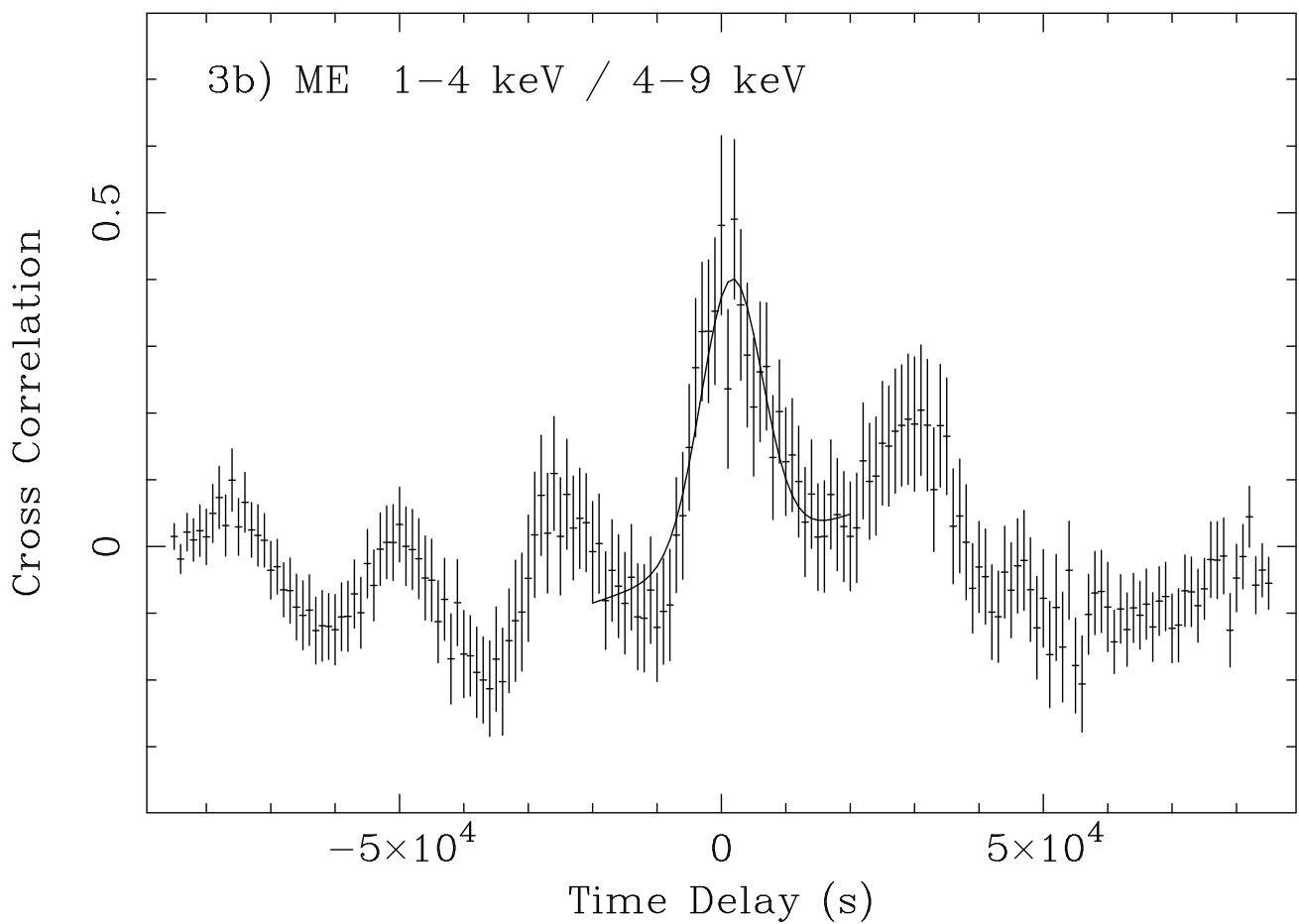
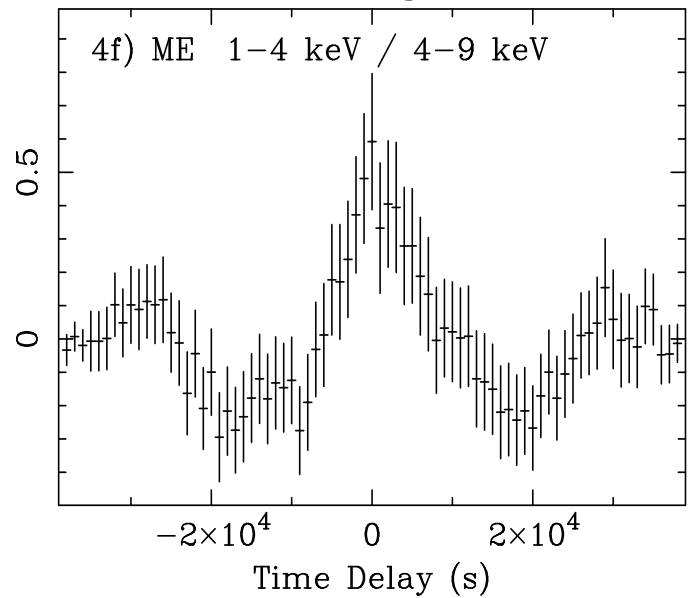
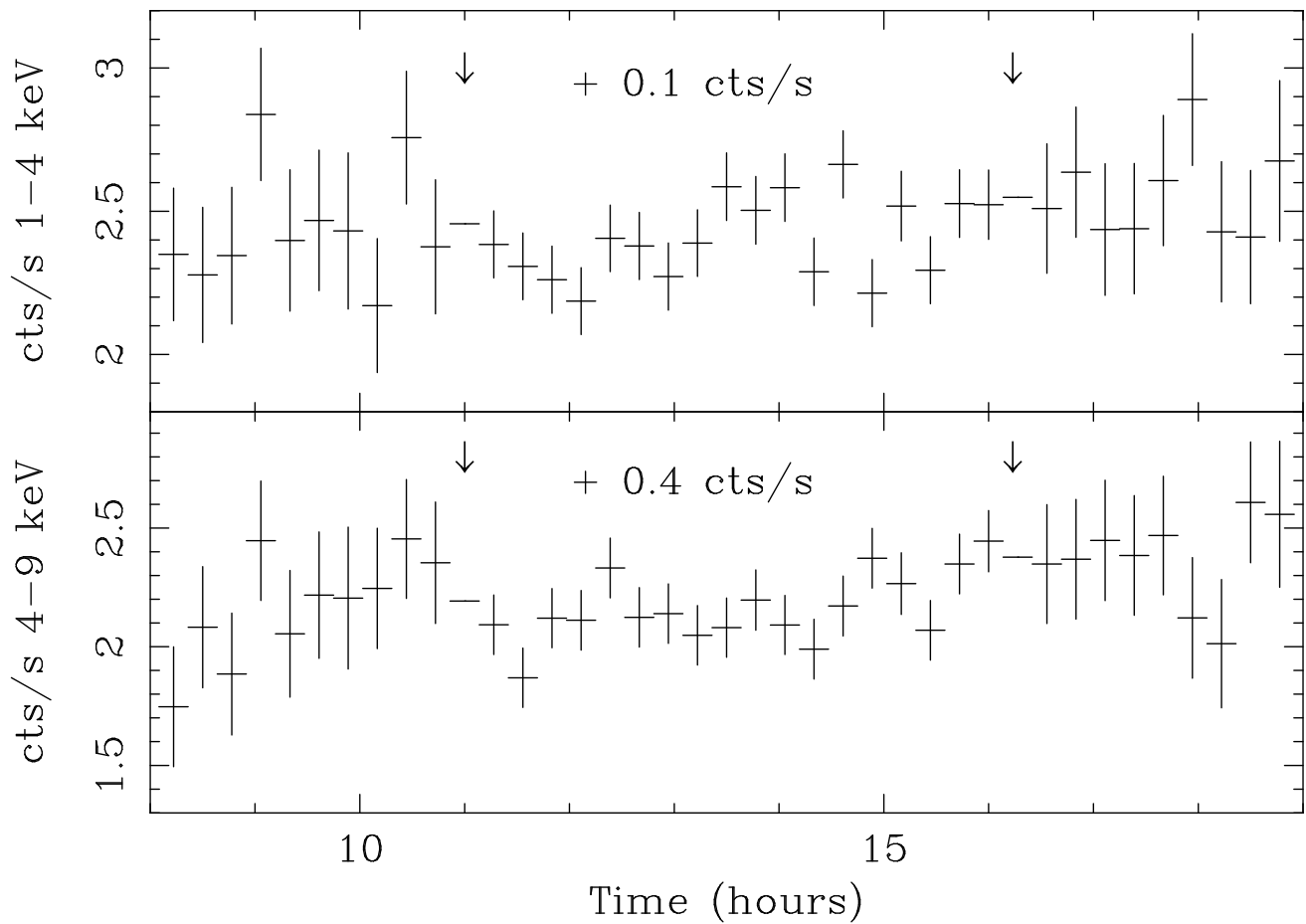
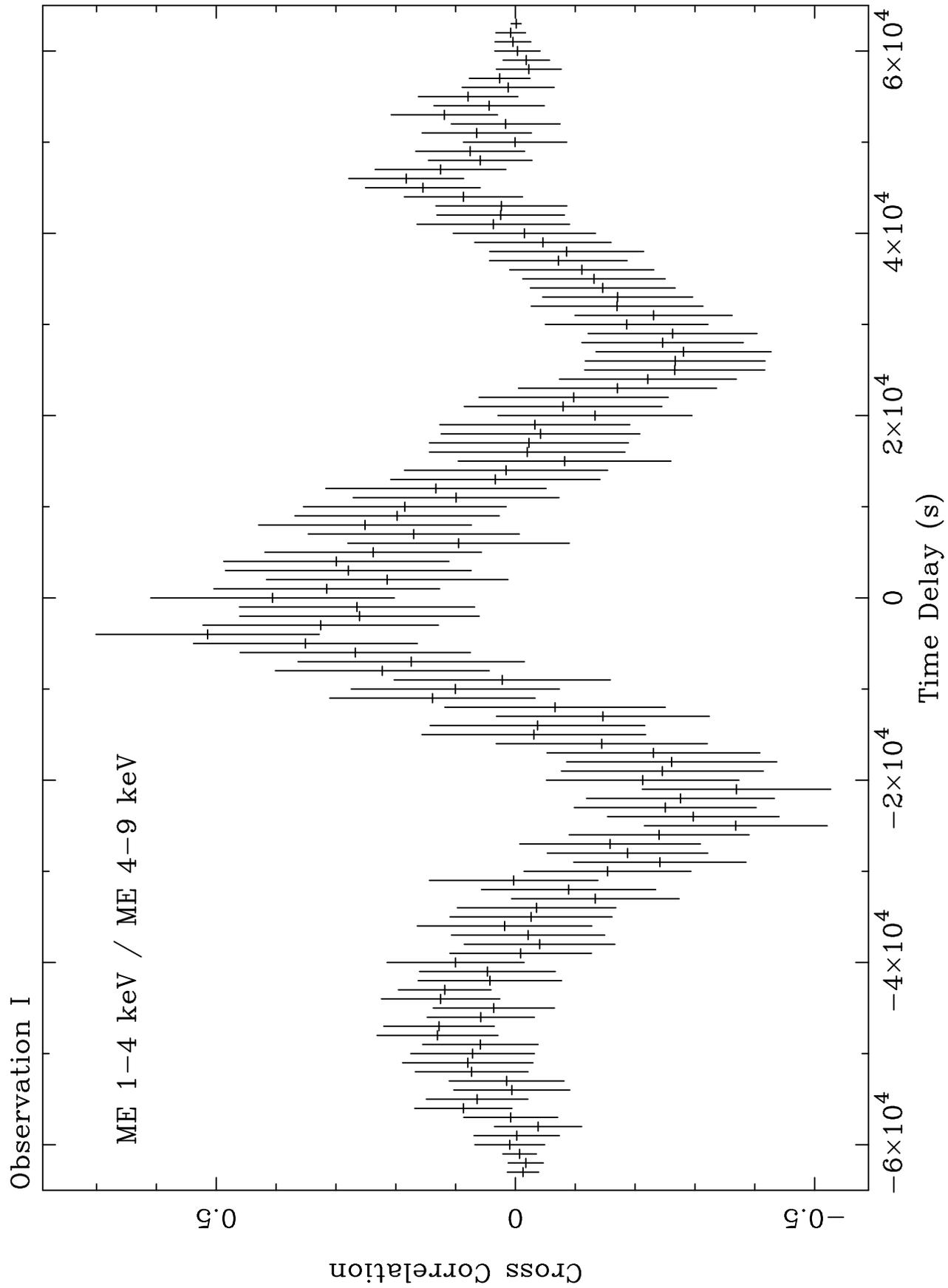
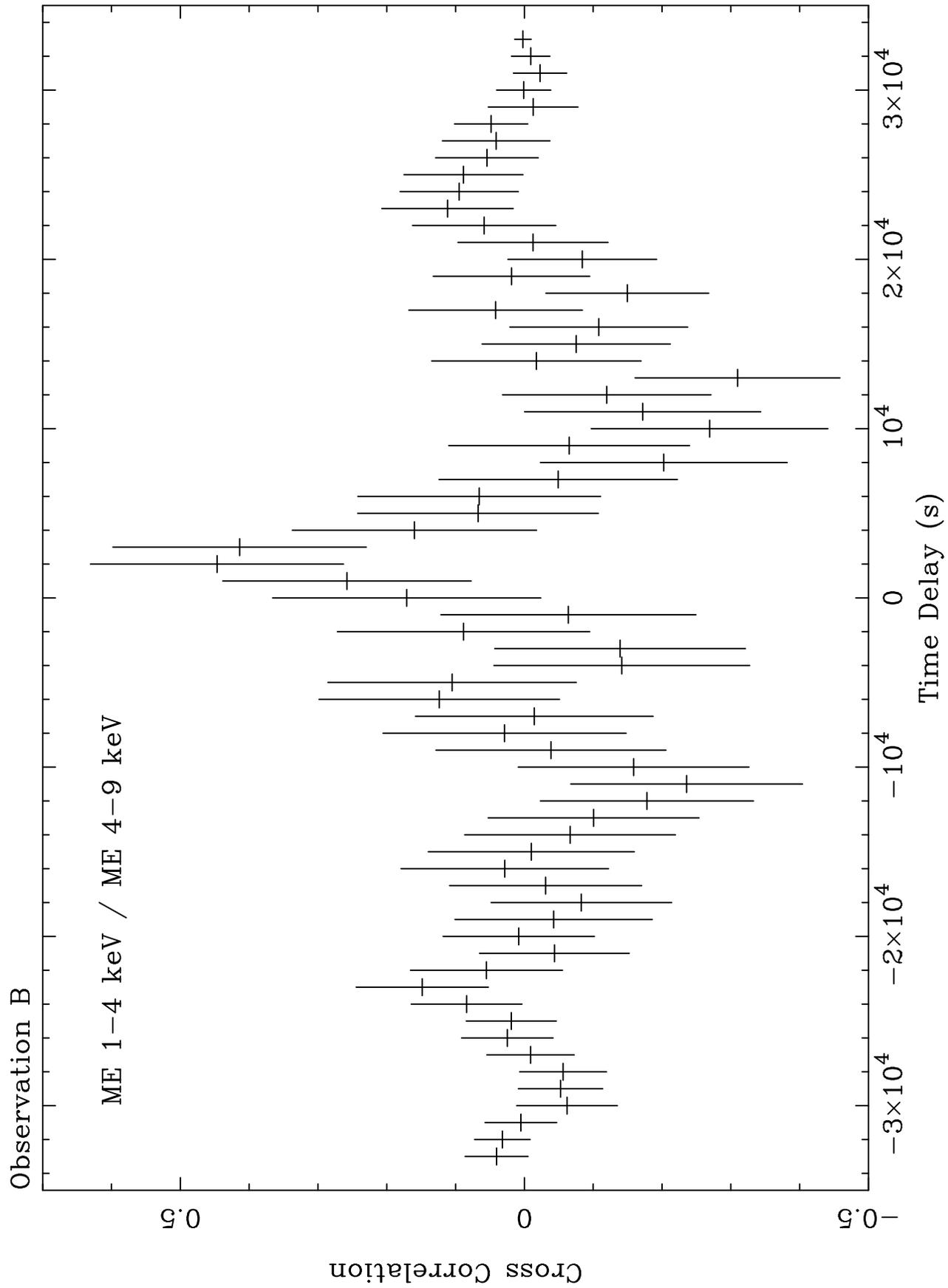
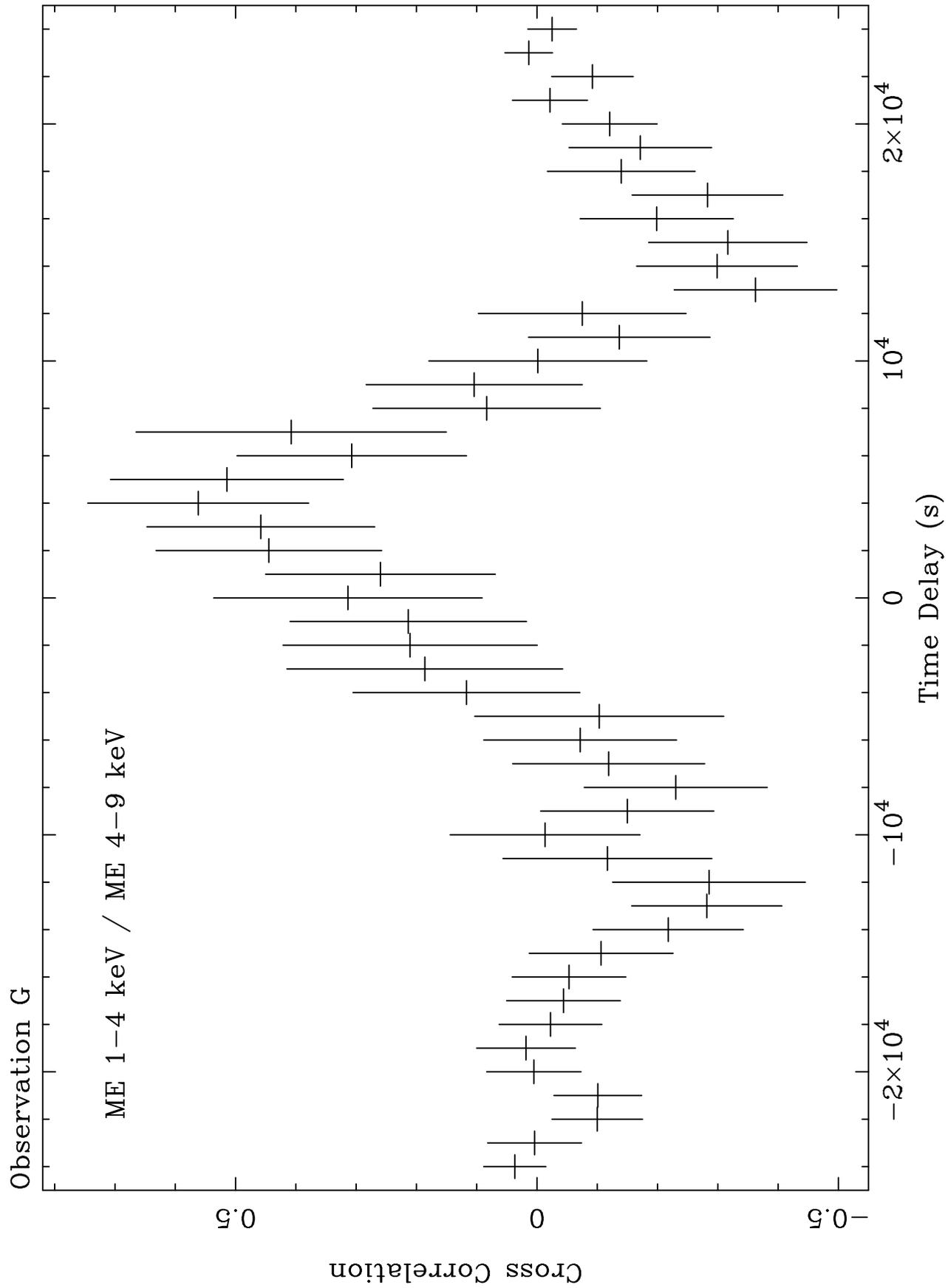
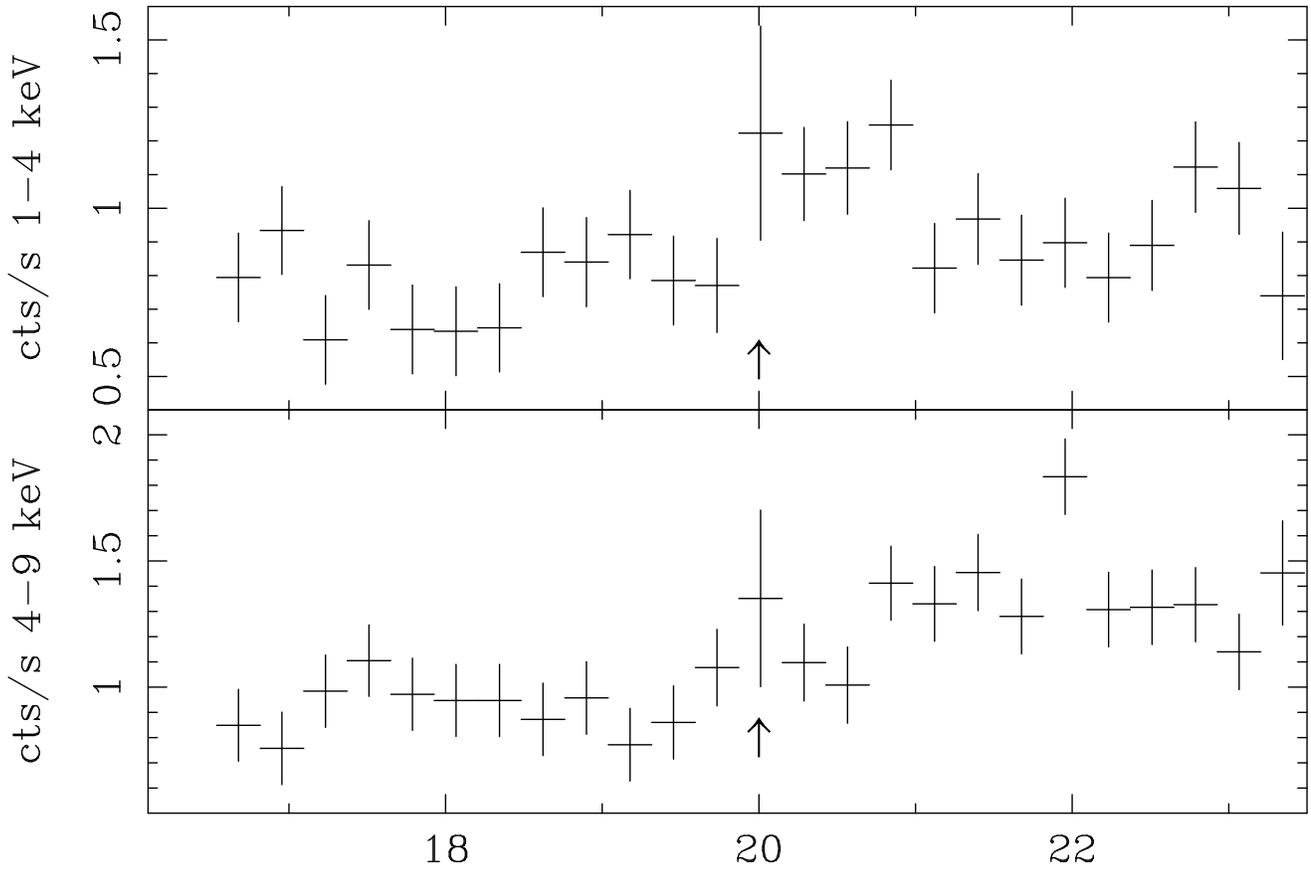
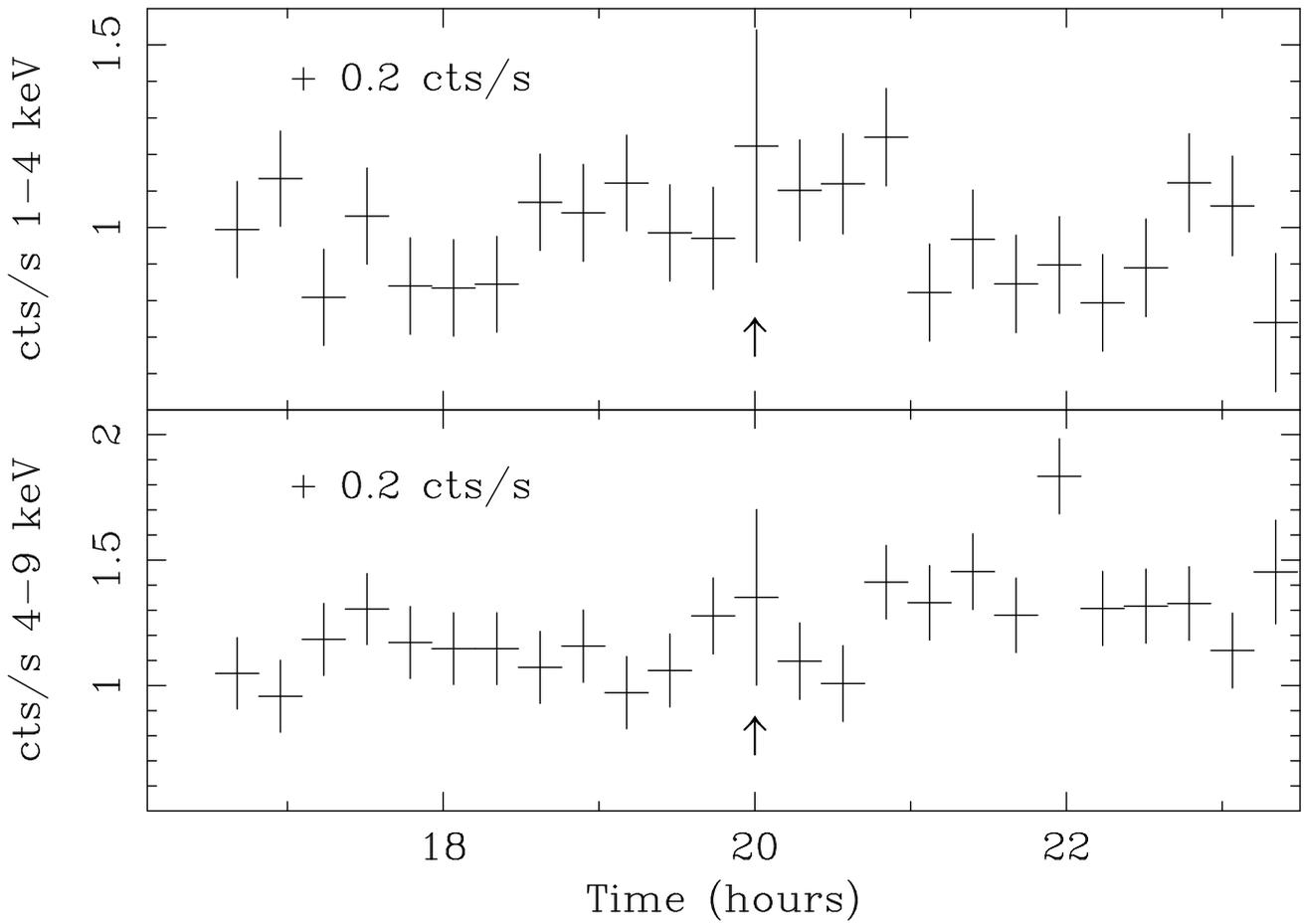
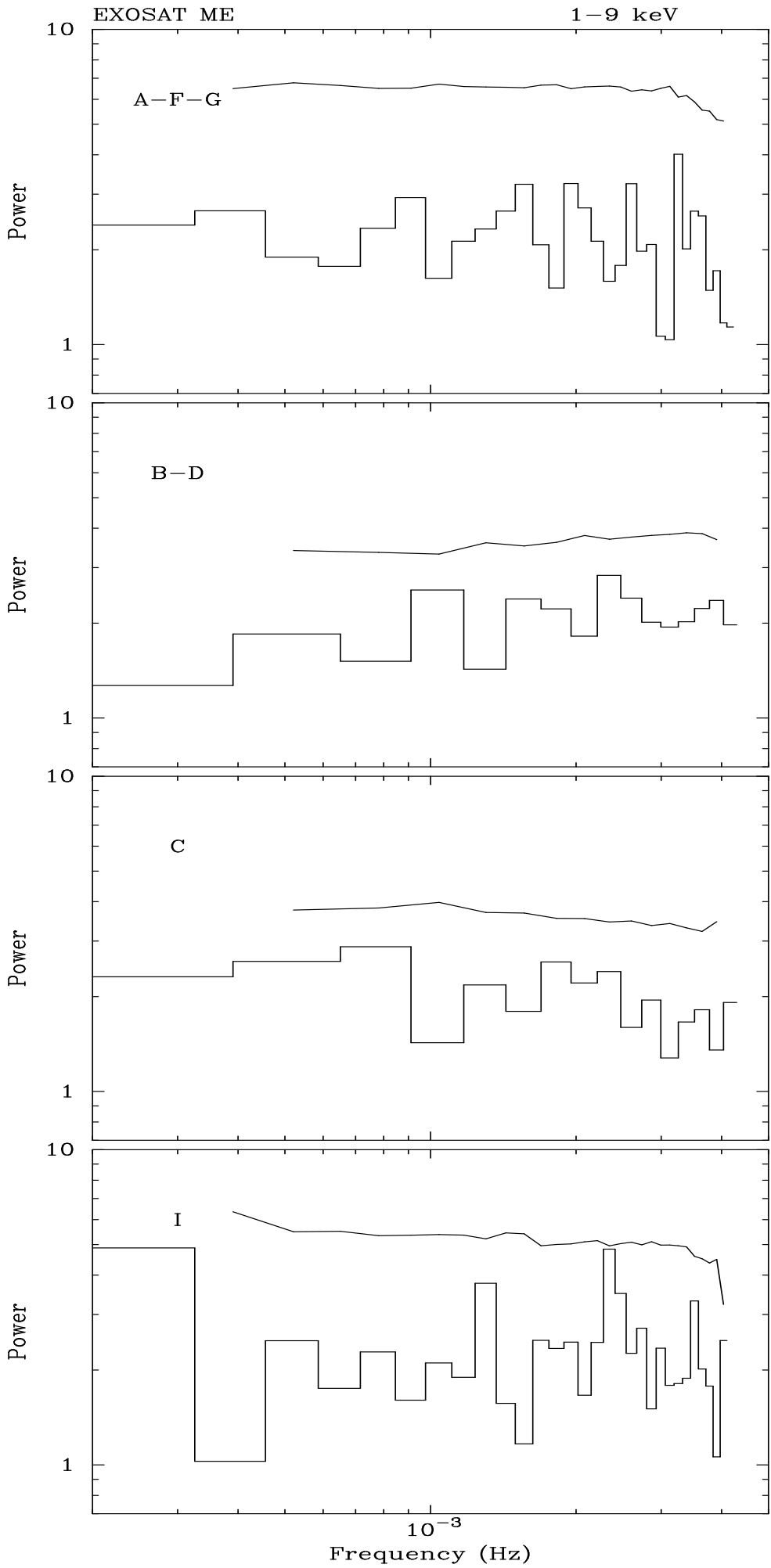
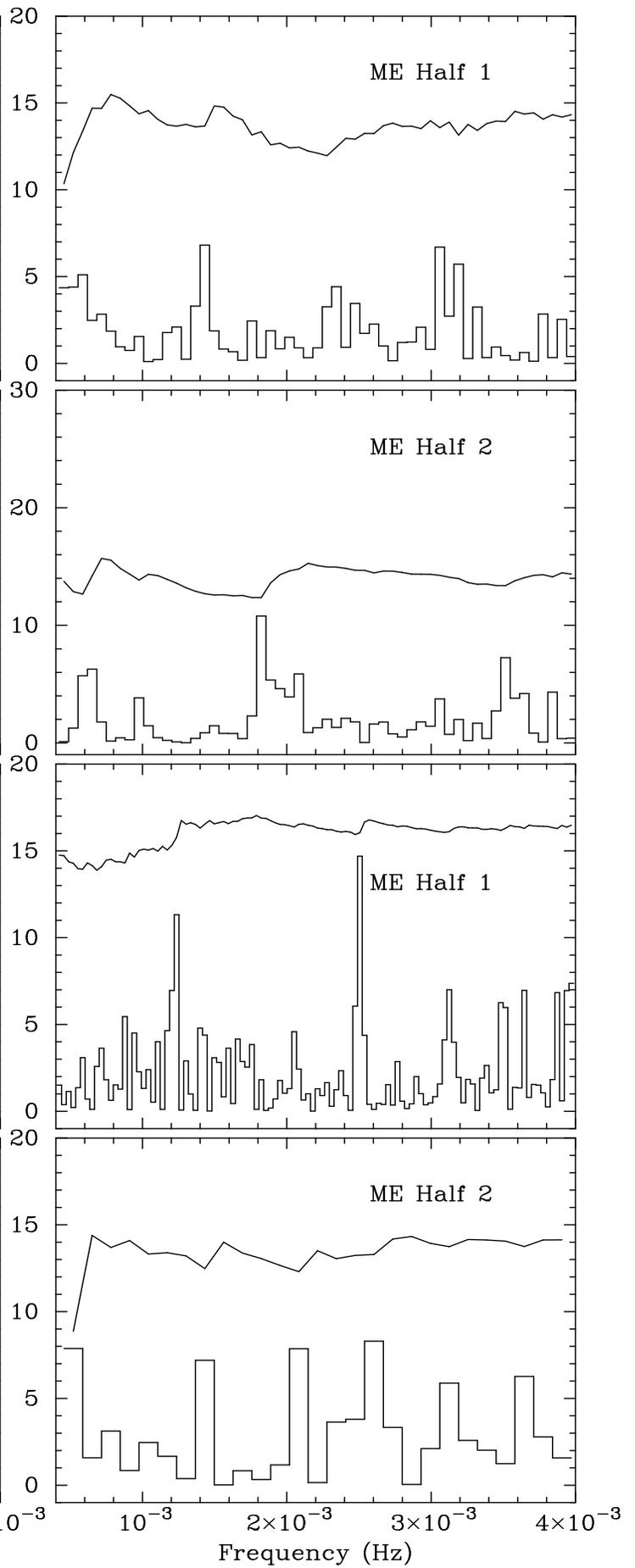
\end{document}